\documentclass[pra,superscriptaddress,nofootinbib]{revtex4-1}

\usepackage[dvips]{graphicx} 
\usepackage{amsfonts}
\usepackage{amssymb}
\usepackage{amscd}
\usepackage{amsmath}    
\usepackage{amsthm}   
\usepackage{enumerate}
\usepackage{epsfig}
\usepackage{subfigure}
\usepackage{url}
\usepackage{multirow}

\newtheorem{theorem}{Theorem}[section]
\newtheorem{lemma}[theorem]{Lemma}

\newtheorem{definition}[theorem]{Definition}

\begin{document}

\title{Postprocessing for quantum random-number generators: entropy evaluation and randomness extraction}

\author{Xiongfeng Ma}
\email{xma@tsinghua.edu.cn}
\affiliation{Center for Quantum Information, Institute for Interdisciplinary Information Sciences, Tsinghua University, Beijing, China}
\affiliation{%
Center for Quantum Information and Quantum Control,\\
Department of Physics and Department of Electrical \& Computer Engineering,\\
University of Toronto, Toronto,  Ontario, Canada}

\author{Feihu Xu}
\email{feihu.xu@utoronto.ca}

\author{He Xu}

\affiliation{%
Center for Quantum Information and Quantum Control,\\
Department of Physics and Department of Electrical \& Computer Engineering,\\
University of Toronto, Toronto,  Ontario, Canada}

\author{Xiaoqing Tan}
\email{ttanxq@jnu.edu.cn}
\affiliation{%
Department of  Mathematics, College of Information Science and Technology, \\
Jinan University, Guangzhou, Guangdong, China}
\affiliation{%
Center for Quantum Information and Quantum Control,\\
Department of Physics and Department of Electrical \& Computer Engineering,\\
University of Toronto, Toronto,  Ontario, Canada}

\author{Bing Qi}
\email{bqi@physics.utoronto.ca}

\author{Hoi-Kwong Lo}
\email{hklo@comm.utoronto.ca}

\affiliation{%
Center for Quantum Information and Quantum Control,\\
Department of Physics and Department of Electrical \& Computer Engineering,\\
University of Toronto, Toronto,  Ontario, Canada}

\begin{abstract}
Quantum random-number generators (QRNGs) can offer a means to generate information-theoretically provable random numbers, in principle. In practice, unfortunately, the quantum randomness is inevitably mixed with classical randomness due to classical noises. To distill this quantum randomness, one needs to quantify the randomness of the source and apply a randomness extractor. Here, we propose a generic framework for evaluating quantum randomness of real-life QRNGs by min-entropy, and apply it to two different existing quantum random-number systems in the literature. Moreover, we provide a guideline of QRNG data postprocessing for which we implement two information-theoretically provable randomness extractors: Toeplitz-hashing extractor and Trevisan's extractor.
\end{abstract}

\maketitle

\section{Introduction}
Random numbers play a crucial role in many fields of science, technology, and industry---for instance, cryptography, statistics, scientific simulations \cite{Meteopolis:Monte:1949}, and lottery. Pseudorandom-number generators (pseudo-RNGs) based on computational complexities have been well developed in the past few decades \cite{Nisan:Design:1994} and can generate high-speed random numbers with little cost. However, the main drawback of pseudo-RNGs is that the generated randomness is not information-theoretically provable. In fact, all of the (software-based) pseudo-RNGs can be realized by a deterministic algorithm given sufficient computational power. This pseudorandomness would cause problems in many applications, such as those in cryptography \cite{Bennett:BB84:1984,Schneier:Crypto:1995}. Recently, Microsoft confirms that XP contains RNG bugs;\footnote{See the news from the Computerworld, ``Microsoft confirms that XP contains random number generator bug,'' Gregg Keizer, November 21, 2007.} security flaws have been found in online encryption methods due to imperfections of random-number generation.\footnote{See the news from the New York Times, ``Flaw Found in an Online Encryption Method,'' John Markoff, February 14, 2012.}

To address the security issues introduced by pseudo-RNGs, physical RNGs have been developed \cite{Jennewein:QRNG:2000,Uchida:RNG:2008,Reidler:RNG:2009}. In particular, the probabilistic nature of quantum mechanics offers a natural way to build an information-theoretically provable RNG \cite{Jennewein:QRNG:2000}---quantum random-number generators (QRNGs). Note that some physical RNGs have been included in microprocessors,\footnote{See, for example, ``Intel Corporation Intel 810 Chipset Design Guide,'' June 1999, Ch.~1.3.5, pages 1--10, download.intel.com/design/chipsets/designex/29065701.pdf.}
although the generated randomness is not quantum mechanical in nature.\footnote{``Evaluation of VIA C3 `Nehemiah' Random Number Generator,'' Cryptography Research, Inc., February 2003, www.cryptography.com/public/pdf/VIA\_rng.pdf.}

In theory, a QRNG can produce random numbers with provable randomness. In practice, quantum signals (the source of true randomness \footnote{We define ``true randomness'' when the randomness is information-theoretically provable. One of the main objectives of our work is to give a randomness extraction procedure such that the extracted output numbers are proven random. On one hand, quantum mechanics comes with randomness in nature. For example, no one can predict the results of the vacuum fluctuation. On the other hand, it is not clear whether or not one can extract ``true'' randomness from the classical noises. That is, we have not proved the randomness extracted from the classical noises. Thus, from a conservative point of view (especially for the usage in cryptography), we only extract random numbers that are information-theoretically provable.}) are inevitably mixed with classical noises. An adversary (Eve) can, in principle, control the classical noise and gain partial information about the raw random numbers. In this work, we assume a trusted device scenario, but the classical noise might be deterministic if we calibrate the system more carefully. For instance, imagine that an input to our device is an \emph{external} power supply. Imagine further that through carefully monitoring the input value of the power to our device, we have determined that the power may still fluctuate by, say, up to 1\%. In principle, the source of such fluctuations of an external power supply might be the action of an adversary---Eve---who, therefore, has complete knowledge about the actual value of the power supply at all times.

Therefore, it is necessary to apply a postprocessing procedure to distill out the true randomness that Eve has \emph{almost} no information about. This distilling procedure is called \emph{randomness extraction}, realized by employing \emph{randomness extractors}. In other words, randomness extractors are used for distilling the true randomness and eliminating the effect of classical noises. The goal of randomness extractors \cite{Impagliazzo:Leftover:1989} is  to extract (almost) perfect randomness from the raw data generated from a practical QRNG with the help of a short random seed, which requires an extra source of randomness. The key input parameter of a randomness extractor is the min-entropy (see Definition \ref{Def:Ext:MinEntropy}) of raw data. Nonetheless, a general method to quantify the min-entropy from the raw data of a QRNG is still missing.

For randomness extraction, previously, some simple methods have been widely used for QRNGs. For instance, an exclusive-\textsc{or} (\textsc{xor}) operation has been employed in the literature \cite{Epstein:RNGXOR:2003,Qi:QRNG:2010}: dividing the raw data into two bit strings and performing a bitwise \textsc{xor} operation between them. In addition, a least-significant-bits operation \cite{Uchida:RNG:2008,Reidler:RNG:2009} or nonuniversal hashing functions \cite{Wayne:QRNG:2010} have been proposed and implemented for QRNGs. These operations can certainly refine the raw data to pass some randomness statistical tests. However, the key point is, the generated randomness is not information-theoretically provable. Recently, a more sophisticated randomness extraction procedure is proposed in Ref.~\cite{Gabriel:QRNG:2010}, which quantifies the randomness by Shannon entropy instead of min-entropy and applies non-universal hash functions for extraction. Unfortunately, the randomness extracted there is still not information-theoretically provable due to the following two reasons: randomness cannot be well quantified by Shannon entropy \cite{Chor:Random:1985,Zuckerman:Random:1990} and the randomness from nonuniversal hashing functions relies on computational assumptions.

In contrast, an important and promising randomness extractor, Trevisan's extractor~\cite{Trevisan:Extractor:1999,Raz:Extractor:1999}, raised considerable theoretical interest not only because of its data parsimony compared to other constructions, but particularly because it is secure against quantum adversaries~\cite{De:ExtQuant:2009}. The seed length of Trevisan's extractor is polylogarithmic in the length of the input and it can also be proven to be a strong extractor (see Theorem 22 in \cite{Raz:Extractor:2002}). That is, the random seed of Trevisan's extractor can be reused. This is particularly important since for a popular universal-hashing function such as Toeplitz hashing \cite{Mansour:Hash:2002,Krawczyk:Hash:94}, which has been well developed for privacy amplification \cite{Bennett:PA:1995} in quantum key distribution (QKD)\footnote{Randomness extractors can also be used for privacy amplification \cite{Bennett:PA:1995} in quantum key distribution (QKD). Note that privacy amplification is a crucial step in QKD postprocessing. A few randomness extractors have been proven to be secure against quantum side channels \cite{De:ExtQuant:2009}. The main advantage is that no (or little) classical communication is required for privacy amplification. It is an interesting prospective research topic to apply the techniques developed in randomness extraction for privacy amplification.}---the random seed used to construct a Toeplitz matrix is longer than the output string. This means that \emph{no} net randomness can be extracted if the universal hashing is directly used for randomness extraction. Despite the considerable theoretical attention of Trevisan's extractor, its practical implementation is still missing. This is probably because Trevisan's extractor has a rather complex structure that the quantum information community may not be familiar with and its implementation involves the non-trivial tradeoff between speed and the values of security parameters.

Here, we fill the above two gaps: a general method for the quantification of randomness and a practical implementation of Trevisan's extractor. We report a generic scheme that can process the raw data from a QRNG to random numbers that (nearly) follow a uniform distribution. The two main contributions of this work can be stated as follows.
\begin{enumerate}
\item
We present a framework for quantifying the quantum randomness from QRNG by min-entropy and discuss how one can evaluate this min-entropy in a physical device. We apply our framework to two different existing QRNGs in the literature \cite{Xu:QRNG:2012,Gabriel:QRNG:2010}.

\item
We provide prototypical implementation of Trevisan's extractor and show how such implementation can be used in real-life QRNGs. From this implementation, we discover the major computational step that limits Trevisan's extractor speed.
\end{enumerate}

Based on a few reasonable assumptions on the physical model of QRNG, the randomness extracted from the proposed postprocessing is information-theoretically provable. Our generic postprocessing scheme consists of three steps: (a) model and characterize the QRNG setup through measurements; (b) quantify the quantum randomness of the raw data with min-entropy; and (c) apply a randomness extractor. To illustrate the generality of our method, our scheme is applied to \emph{two} different types of real-life QRNGs \cite{Xu:QRNG:2012, Gabriel:QRNG:2010}.

Besides attempt of Trevisan's extractor, we also implement the Toeplitz-hashing extractor and prove that the universal-hashing function~\cite{WC_Authen_79} constructs a strong extractor (see Definition \ref{Def:Ext:StrongExt}), and thus allows us to maintain the randomness of the seed for subsequent applications. The implementation speeds for Trevisan's and the Toeplitz-hashing extractors are 0.7 and 441 kb/s, respectively. The outcomes from both extractors pass all the standard randomness tests we exploited.

Compared to the Toeplitz-hashing, Trevisan's extractor requires a seed with a shorter length. Thus, Trevisan's extractor has a certain advantage in cases where the random seed bits are limited. We note that our prototypical implementation of Trevisan's extractor allows researchers to better understand the complexity and difficulty in the implementation of Trevisan's extractor, thus paving the way to future implementations. After the completion and posting of a preliminary version of our paper on arXiv~\cite{MXF:Extractor:2012}, Maurer \emph{et al.}~posted a follow-up paper on our work in which they  built on our results and improved the speed of implementation \cite{Mauerer:Extraction:2012}.


The rest of this paper is organized as follows. We introduce related notations and definitions in the rest of this section. In Sec.~\ref{Sec:Ext:RandomEva}, we present a procedure to evaluate the min-entropy of the quantum signals. We implement Trevisan's extractor in Sec.~\ref{Sec:Ext:Trevisan} and a universal hashing (Toeplitz-hashing) extractor in Sec.~\ref{Sec:Ext:Hash}. In Sec.~\ref{Sec:Ext:Test}, we show the results of statistical tests. Finally, we conclude this paper in Sec.~\ref{Sec:Eff:Conclusion}.

\subsection{Notations and definitions}
\emph{Notations:} $U_d$ represents a uniform distribution on $\{0,1\}^d$; $\log$ denotes the logarithm to base 2. The outcome of an \emph{ideal} RNG, described by a random variable, follows a uniform distribution.

Min-entropy is widely used for quantifying the randomness of a probability distribution \cite{Chor:Random:1985,Zuckerman:Random:1990}.
\begin{definition} \label{Def:Ext:MinEntropy}
(Min-entropy). The min-entropy of a probability distribution $X$ on $\{0,1\}^n$ is defined by
\begin{equation} \label{Ext:Def:minentropy}
\begin{aligned}
H_\infty(X) = - \log\left( \max_{v\in\{0,1\}^n}\mathrm{Prob}[X=v] \right).
\end{aligned}
\end{equation}
\end{definition}

In cryptography, the deviation of a practical protocol from an ideal protocol is characterized by a \emph{security parameter}, $\varepsilon$. The statistical distance is commonly used as a standard security measure.
\begin{definition}
($\varepsilon$-close). Two probability distributions $X$ and $Y$ over the same domain $T$ are $\varepsilon$-close if the statistical distance between them is bounded by $\varepsilon$,
\begin{equation} \label{Ext:Def:statdistance}
\begin{aligned}
\| X-Y \| &\equiv \max_{V\subseteq T} |\sum_{v\in V}(\mathrm{Prob}[X=v]-\mathrm{Prob}[Y=v])| \\
&= \frac12\sum_{v\in T} |\mathrm{Prob}[X=v]-\mathrm{Prob}[Y=v]| \leq \varepsilon. \\
\end{aligned}
\end{equation}
\end{definition}
Here, the second equality in Eq.~\eqref{Ext:Def:statdistance} can be proven by induction on the domain size $|T|$. It is obvious that the equality holds for $|T|=1,2$. For the case of $|T|\ge3$, one can always find two $v_1$ and $v_2$ in $T$ such that $(\mathrm{Prob}[X=v_1]-\mathrm{Prob}[Y=v_1])(\mathrm{Prob}[X=v_2]-\mathrm{Prob}[Y=v_2])\ge0$. Then, one can combine $v_1$ and $v_2$ together as a new variable $v'$ to form a new domain $T'$. If the statement of the equality $S(|T|)$ holds, the above argument shows that $S(|T|+1)$ also holds. Since the statement holds for $S(1)$ and $S(2)$, by induction, it holds for all $S(|T|)$ with $|T| \geq 1$.

Roughly speaking, the statistical distance quantifies the distinguishability of two probability distributions. The factor of $1/2$ in Eq.~\eqref{Ext:Def:statdistance} is used to normalize the statistical distance so that its value falls into $[0,1]$. When $X$ is $\varepsilon$-close to $Y$, $X$ is indistinguishable from $Y$ except for a small probability $\varepsilon$. For example, the output of a practical RNG is said to be $\varepsilon$-close to an ideal RNG if it satisfies Definition \ref{Ext:Def:statdistance}. We emphasize that the security parameters from Definition \ref{Ext:Def:statdistance} are composable. The notion of composability was first proposed in the classical cryptography for the study of security when composing classical cryptographic protocols in a complex manner~\cite{Canetti2001,Canetti2002}. It is introduced to quantum cryptography by Refs.~\cite{BenOr:Security:05,Renner:Security:05}.

\begin{definition} \label{Def:Ext:Extractor}
(Extractor). A ($k$,$\varepsilon$,$n$,$d$,$m$)-extractor is a function
\begin{equation} \label{Ext:Def:extractor}
\begin{aligned}
\mathrm{Ext}: \{0,1\}^n\times\{0,1\}^d\rightarrow\{0,1\}^m,
\end{aligned}
\end{equation}
such that for every probability distribution $X$ on $\{0,1\}^n$ with $H_\infty(X)\ge k$, the probability distribution $\mathrm{Ext}(X,U_d)$ is $\varepsilon$-close to the uniform distribution on $\{0,1\}^m$.
\end{definition}
In short, an extractor is a function that takes a small seed of $d$ bits and a partially random source of $n$ bits to output an almost perfect random bit string of $m$ bits.

\begin{definition} \label{Def:Ext:StrongExt}
(Strong extractor). A ($k$,$\varepsilon$,$n$,$d$,$m$)-strong extractor $\mathrm{Ext}(X,U_d)$ is an extractor such that the probability distribution $\mathrm{Ext}(X,U_d)\circ U_d$ is $\varepsilon$-close to the uniform distribution on $\{0,1\}^{m+d}$.
\end{definition}
Note that the key advantage of a strong extractor is that the input (random) seed can be reused (with a security parameter increased by $\varepsilon$). Thus, one can partition the output of a practical RNG into (small) blocks and process them by a strong extractor with the same seed.

\begin{definition} \label{Def:Ext:UniversalHash}
(Universal hashing). A family of hash functions $\mathcal{H}$, mapping $S$ to $T$, is two-universal if
\begin{align}
\label{eqn-2universal-def1}
\mathrm{Prob}_{h\in\mathcal{H}} \{ h(x) = h(y)\} \leq \frac{1}{|T|},
\end{align}
for all  $x \neq y \in S$.
\end{definition}

\section{Quantum randomness evaluation} \label{Sec:Ext:RandomEva}
In this section, we provide a general framework to evaluate the quantum (true) randomness from a practical QRNG. The QRNGs developed in Refs.~\cite{Xu:QRNG:2012, Gabriel:QRNG:2010} are discussed as illustrations of the evaluation process. We note that our evaluation procedure is a generic method that can be applied to other QRNGs with certain modifications.

\subsection{Physical model} \label{Sub:Ext:Model}
In general, the random numbers of a QRNG come from a certain measurement. We refer to the measurement outcome as \emph{quantum signal}. This quantum signal is inevitably mixed with \emph{classical noises}, such as background detections and electronic noises. From a cryptographic view, these classical noises might be known to (or even manipulated by) Eve in the worst case scenario. Hence, the main objective of the postprocessing for a QRNG is to extract out the quantum (true) randomness and eliminate the contributions of classical noises.

\begin{figure}[tbh]
\centering \resizebox{12cm}{!}{\includegraphics{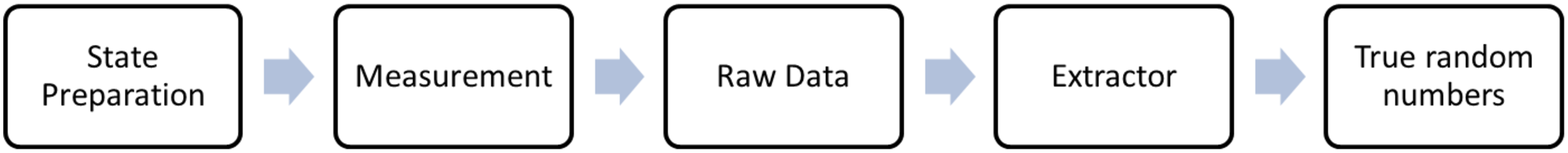}}
\caption{A generic schematic diagram of a QRNG setup. A quantum state is prepared and measured. Then the outcome is processed to generate random numbers by an extractor.} \label{Fig:Ext:Flow}
\end{figure}

Let us consider a generic flow chart of QRNG, as shown in Fig.~\ref{Fig:Ext:Flow}. First, a quantum state is prepared, which is the source of true randomness. Then, a measurement is performed on the quantum state. Finally, the raw data is postprocessed by a randomness extractor to generate truly random numbers. For example, the quantum state in Ref.~\cite{Xu:QRNG:2012}, which characterizes the random phases of the photons from spontaneous emissions, is prepared by operating a laser near its threshold level; the measurement is operated by a delayed self-heterodyning system; and the raw data is evaluated based on a physical model and processed by randomness extractors. For the QRNG in Ref.~\cite{Gabriel:QRNG:2010}, the quantum state is produced by the quadrature amplitude of the vacuum state; the measurement is realized by a homodyne detector; and the raw data is processed by a hash extractor.

\subsection{Quantum randomness evaluation} \label{Sub:Ext:RandEva}
The key parameter we need to evaluate here is the min-entropy, defined in Eq.~\eqref{Ext:Def:minentropy}, of the quantum signal contained in the raw data. In the following, we present a method to evaluate the min-entropy by deriving the probability distribution of the quantum signal.

Let us take the QRNG setup in Refs.~\cite{Qi:QRNG:2010,Xu:QRNG:2012} as the first example to show the detailed procedures of the evaluation process. In this case, the quantum signal comes from vacuum fluctuation and the contribution from all other sources of phase fluctuations is defined as classical noise. The details of this QRNG can be found in Ref.~\cite{Xu:QRNG:2012}. The assumptions of the physical model needed for the derivation of the probability distribution of the quantum signal are listed as follows:
\begin{enumerate}
\item
The total signal is a mixture of quantum signal and classical noise. Quantum signal is independent of classical noise.

\item
The quantum signal follows a Gaussian distribution. The total analog signal is digitalized by an analog-to-digital convertor (ADC).

\item
The ratio between the variances of quantum signal and classical noise can be determined, denoted by $\gamma$.

\item
Total signal variance can be characterized by sampling, denoted by $\sigma^2_{total}$.
\end{enumerate}
Note that the last assumption can be satisfied when the sequence of the raw data is independent and identically distributed (i.i.d.). Also, there is no need to assume that classical noise follows a Gaussian distribution.

In summary, to derive the probability distribution of the quantum signal, the key point is to find out its variance. This is done by measuring the total variance of the raw data and the quantum-to-classical ratio. That is, with assumptions 1, 3 and 4, one can easily derive the quantum variance,
\begin{equation} \label{Extractor:RandomEva:Qvar}
\begin{aligned}
\sigma^2_{quantum} &= \frac{\gamma\sigma^2_{total}}{1+\gamma}. \\
\end{aligned}
\end{equation}
From the variance together with the Gaussian distribution assumption, one can get the whole probability distribution of the quantum signal. Since the analog signal is sampled by an 8-bit ADC to generate digital bits \cite{Xu:QRNG:2012}, one can evaluate the probability distribution of the digitalized output on $\{0,1\}^8$, given the Gaussian distribution of the quantum signal. Then the min-entropy of the quantum signal can be derived by Definition \ref{Def:Ext:MinEntropy}. Following the detailed calculation procedures in Ref.~\cite{Xu:QRNG:2012}, a min-entropy of 6.7 bits per 8-bit raw sample (from an 8-bit ADC) is obtained. In Ref.~\cite{Xu:QRNG:2012}, the authors use the ADC to cut off bins evenly along the voltage axis. By carefully designing the bin size, one might increase the min-entropy.

Next, we discuss how our evaluation process can be applied to Ref.~\cite{Gabriel:QRNG:2010}. The assumptions of a physical model are similar to the ones discuss above in that the vacuum state fluctuation used there~\cite{Gabriel:QRNG:2010} also follows a Gaussian distribution. To rigorously quantify the randomness by min-entropy, from the data about the variance of total signal and classical noise there (see Fig.~2 in Ref.~\cite{Gabriel:QRNG:2010}), the variance of quantum signal can be derived. Afterwards, instead of calculating Shannon entropy there, the min-entropy of the quantum signal can be derived by Definition~\ref{Def:Ext:MinEntropy} given the Gaussian distribution of the quantum signal about the probability distribution of the digitalized output on $\{0,1\}^{16}$.\footnote{Without the detailed data values of Ref.~\cite{Gabriel:QRNG:2010}, we did not calculate the final result of min-entropy. As indicated earlier in Sec.~\ref{Sub:Ext:RandEva}, the missing experimental parameters are the total variance of quantum signals and classical noises and the ratio between the variances of quantum signals and classical noises.}

We note that our framework can also be applied to other types of QRNGs, such as those with discrete variables rather than continuous variables. The key point is to evaluate the min-entropy of quantum signals through quantitatively separating its contributions from quantum signals and classical noises. Such an extension would be an interesting prospective research topic.

\subsection{Upper bound of randomness}
The randomness of a given QRNG setup is a limited resource. This can be shown by providing the upper bound of randomness, say, via Shannon entropy, that one can extract from the measurement outcome.\footnote{Roughly speaking, the min-entropy can be regarded as the lower bound of randomness one can extract, whereas Shannon entropy can be treated as the upper bound. The min-entropy is always no greater than the corresponding Shannon entropy.} The upper bound also indicates how much margin is left for further improvement in postprocessing.

Here, since we only have the experimental data of Ref.~\cite{Xu:QRNG:2012}, we take it as an example to show how one can evaluate the upper bound of entropy for a practical QRNG setup. The quantum signal is measured by a photo detector (PD). Given a perfect photon-number resolving detector, the upper bound of the min-entropy is determined by the photon number within the detection time window. The laser power used in the setup is 0.95 mW, \footnote{In principle, one can go beyond this limitation by increasing the laser power. However, the upper bound of min-entropy can only increase logarithmically with power intensity.} which corresponds to $1.5\times10^6$ photons at 1550 nm within 200 ps detection time window. Thus, the maximal entropy of a sample from the PD can be estimated by $\log(1.5\times10^6)=20.5$ bits, which is the upper bound of the min-entropy of the QRNG source.

\section{Randomness Extraction} \label{Sec:Ext:Extration}
In this section, we will present our prototypical implementations of Trevisan's extractor and the Toeplitz-hashing extractor. These implementations can be easily used for the extraction of Refs.~\cite{Xu:QRNG:2012, Gabriel:QRNG:2010}, and other more general QRNGs.

\subsection{Trevisan's extractor} \label{Sec:Ext:Trevisan}

\subsubsection{Results summary}
Trevisan proposed an approach to construct randomness extractors
based on pseudorandom-number generators~\cite{Trevisan:Extractor:1999}. Trevisan's extractor has a number of important theoretical advantages.\footnote{For a discussion, see, for example, a follow-up paper by Mauerer \emph{et al.} \cite{Mauerer:Extraction:2012}.} First, it is secure against a quantum adversary.  Second, the seed length is polylogarithmic in the length of the input. Third, it can also be proven to be a strong extractor with certain modifications on the security parameters (Theorem 22 in \cite{Raz:Extractor:2002}). However, a real-life implementation of this important extractor was never reported in the literature.


Here, we implement its improved version by Raz, Reingold and Vadhan \cite{Raz:Extractor:1999}. There are two main steps to construct Trevisan's extractor: one-bit extractor and combinatorial design. The one-bit extractor can be realized by an error correcting code, which is constructed by concatenating a Reed-Solomon code with a Hadamard code, as shown in Appendix A of Ref.~\cite{Raz:Extractor:2002}. For the combinatorial design part, we implement a refined version of Nisan-Wigderson design \cite{Nisan:Design:1994,MXF:Design:2011}.

Our implementation of Trevisan's extractor yields an output speed of 0.7 kb/s. From our result, we locate the major computational step that limits Trevisan's extractor speed, which lies on the error-correction based one-bit extractor. After the completion and posting of a preliminary version of our paper on arXiv~\cite{MXF:Extractor:2012}, Maurer et al.~posted a follow-up paper \cite{Mauerer:Extraction:2012} on our work, where alternative one-bit extractors, rather than listing error correcting codes, are utilized. As a result, the extraction speed in the implementation of Maurer \emph{et al.} is substantially improved \cite{Mauerer:Extraction:2012}, which can go up to 150 kb/s. Compared to the Toeplitz-hashing extractor (Sec.~\ref{Sec:Ext:Hash}), Trevisan's extractor requires a seed with a shorter length. Thus, Trevisan's extractor has a certain advantage in cases where the random seed bits are limited. Although our implementation results show that the speed of the Toeplitz-hashing extractor (441 kb/s) is much faster than that of Trevisan's extractor (0.7 kb/s), a detailed comparison between these two extractors is an interesting prospective research project.


\subsubsection{Implementation procedure} \label{Sub:Ext:Together}
We sketch out the implementation procedure in this section, while the implementation details are shown in Appendix~\ref{App:Ext:Trevisan}. We input an $n_i$-bit string, raw data from a QRNG, with a min-entropy at least $k$, and a $d$-bit random seed ($y$) into an $(k,\varepsilon,n_i,d,n_f)$-extractor, constructed by combining an $(n_i,1/2-\varepsilon/4n_f)$-error correcting code\footnote{An $(n,p)$-error correcting code has a codeword length of $n$ and is able to correct error rate up to $p$.} and an $(m_e,\rho)$-design,\footnote{An $(m,\rho)$-design means a collection of $m$ subsets. The average number of overlap elements between two subsets is no more than $\rho$. Details can be found in Ref.~\cite{MXF:Design:2011}.} and then output an $n_f$-bit string which is $\varepsilon$-close to a uniform distribution. Here, $k$, $\varepsilon$ and $n_i$ are given from the source and practical use of random numbers. We need to figure out $d$ (as small as possible) and $n_f$ (as large as possible).

\begin{enumerate}
\item
Map the input $n_i$-bit string to an $\bar{n}$-bit string according to the $(n_i,1/2-\varepsilon/4n_f)$-error correcting code. Here, $\bar{n}$ can be assumed to be a power of 2 \cite{Trevisan:Extractor:1999}. In practice, one can concatenate the Reed-Solomon code and Hadamard code together (see Appendix A of \cite{Raz:Extractor:2002}), where the codeword length is given by
\begin{equation} \label{Ext:design:ECnbar}
\begin{aligned}
\bar{n} &= 2^{2m_e}, \\
m_e &= \lceil \log{n_i}+2\log{n_f}-2\log{\varepsilon} +4 \rceil. \\
\end{aligned}
\end{equation}
Also, $n_f$ can be upper bounded by $k$ for the error correcting code construction.

\item
Construct an $(m_e,\rho)$-design \cite{MXF:Design:2011}, with
\begin{equation} \label{Ext:design:lrho}
\begin{aligned}
m_e &= \frac12\log\bar{n} = O(\log(n_i/\varepsilon)) \\
\rho &= [k-3\log(n_f/\varepsilon)-d-3]/n_f, \\
\end{aligned}
\end{equation}
where the second equation is from Proposition 10 and Theorem 22 in Ref.~\cite{Raz:Extractor:2002}, typically $1\le\rho\le1.5$. The design parameter $\rho$ can be viewed as the ratio of min-entropy that can be extracted. One can simply pick up $\rho=1$ if the output length is to be optimized (Lemma 17 in Ref.~\cite{Raz:Extractor:2002}). The extractor seed, with a length of $d$, is composed of blocks of seeds with lengths of the square of the smallest power of 2 which is greater than $m_e$. Note that this block design idea is proposed by Raz \emph{et al.}~\cite{Raz:Extractor:2002}. Here, we are interested in a design with $\rho=1$, so that most of randomness can be extracted. According to the explicit design proposed by Nisan and Wigderson \cite{Nisan:Design:1994} and proved in Refs.~\cite{Hartman:Design:2003,MXF:Design:2011}, the number of such blocks and hence the seed length are given by
\begin{equation} \label{Ext:design:seedlength}
\begin{aligned}
b &= \lceil\log{n_f}\rceil-m_d+1 \\
d &= 2^{2m_d} b = O(\log^2(n_i/\varepsilon)\log{n_i}) \\
m_d &\equiv \lceil \log{2m_e} \rceil. \\
\end{aligned}
\end{equation}
In fact, any design with $d\ge\lceil \log\bar{n} \rceil^2 b$ and $\rho=1$ can be applied here.

\item
The $i$th bit of the $n_f$-bit output is given by the $y_{S_i}$th bit of the encoded $\bar{n}$-bit string, where $y_{S_i}$ is a substring of $y$, formed by the bits of $y$ at the positions given by the elements of $S_i$.
\end{enumerate}

\subsection{Universal hashing} \label{Sec:Ext:Hash}
Owing to the similarity between the definitions of extractors and privacy amplification \cite{Bennett:PA:1995}, any privacy amplification scheme can be used as an extractor, in principle. However, there is one subtle difference. In privacy amplification, the random seed (public randomness) is assumed to be free, whereas in an extractor, one needs to take the seed into account as it does consume random bits. Therefore, a direct transplant of privacy amplification schemes may not work for randomness extraction. In fact, for a popular universal hashing function, Toeplitz-hashing \cite{Mansour:Hash:2002,Krawczyk:Hash:94}, the random seed used to construct a Toeplitz matrix is longer than the output string. This means that no net randomness can be extracted if the universal hashing is directly used for randomness extraction. To overcome this problem, one needs to prove that the privacy amplification scheme constructs a strong extractor (see Definition \ref{Def:Ext:StrongExt}), thus allowing one to maintain the randomness of the seed for
subsequent applications. Fortunately, the extractors constructed by universal hashing functions \cite{WC_Authen_79} (see Definition \ref{Def:Ext:UniversalHash}) can be easily proven to be strong extractors by the Leftover Hash Lemma
\cite{Impagliazzo:Leftover:1989}.

\begin{lemma} \label{Lemma:Ext:Leftover}
(Leftover Hash Lemma \cite{Impagliazzo:Leftover:1989}). Let $\mathcal{H}=\{h_1,h_2,\dots,h_{2^d}\}$ be a (two-)universal hashing family, mapping from $\{0,1\}^n$ to $\{0,1\}^{m}$, and $X$ be a probability distribution on $\{0,1\}^n$ with $H_\infty(X)\ge k$. Then for $x\in X$ and $h_y\in \mathcal{H}$ where $y\in U_d$, the probability distribution formed by $h_y(x)\circ y$ is $\varepsilon=2^{(m-k)/2}$-close to $U_{m+d}$.
That is, it forms a ($k$,$2^{(m-k)/2}$,$n$,$d$,$m$)-strong extractor.
\end{lemma}

We note that Lemma \ref{Lemma:Ext:Leftover} also implies that the Toeplitz matrix may be re-used in the privacy amplification of QKD. Then, one can use a private key (as a seed) to construct a Toeplitz matrix for privacy amplification without compromising (much of) the privacy of the seed. Hence, reusing the seed can save the classical communication for privacy amplification, which is normally required in standard QKD postprocessing \cite{MXF:Finite:2011}.
It is also practically beneficial for privacy amplification to divide the raw key data into small blocks and apply a small Toeplitz matrix individually. However, the finite-size effect of a small block can significantly lower the privacy amplification efficiency \cite{Fung:Finite:2010}. This issue is an interesting research topic for future study.

Here, we use Toeplitz matrices for universal hashing function construction \cite{Mansour:Hash:2002,Krawczyk:Hash:94} and implement the Toeplitz-hashing extractor. A Toeplitz matrix of dimension $n\times m$ requires only the specification of the first row and the first column, and the other elements of the matrix are determined by descending diagonally down from left to right. Thus, the total number of random bits required to construct (choose) a Toeplitz matrix is $n+m-1$.

The procedure of Toeplitz-hashing extractor is given as follows:
\begin{enumerate}
\item
Given raw data of size $n$ with a min-entropy of $k$ and a security parameter $\varepsilon$, determine the output length to be
\begin{equation} \label{Ext:Universal:ToepOutm}
\begin{aligned}
m = k-2\log\varepsilon. \\
\end{aligned}
\end{equation}

\item
Construct a Toeplitz matrix with an $n+m-1$-random-bit seed.\footnote{For demonstration purposes, we use pseudorandom numbers for this step.}

\item
The extracted random bit string is obtained by multiplying the raw data with the Toeplitz matrix.
\end{enumerate}

We implement a Toeplitz-hashing extractor to the QRNG presented in Ref.~\cite{Xu:QRNG:2012}. As mentioned in Sec.~\ref{Sec:Ext:RandomEva}, the min-entropy of the raw data is bounded by 6.7 bits per sample. With the input bit-string length of $2^{12}=4096$, the output bit-string length is $4096\times6.7/8\ge3430$. Thus, we use a $4096\times3230$ Toeplitz matrix for randomness extraction, which results in $\varepsilon<2^{-100}$ as from Eq.~\eqref{Ext:Universal:ToepOutm}. Our implementation of a Toeplitz-hashing extractor achieves generation rates of 441 kbits/s.\footnote{Toeplitz hashing can be implemented much faster with hardware implementation \cite{Krawczyk:Hash:94}.}


Notice that recently Toeplitz matrix hashing is implemented for QKD privacy amplification with a block size exceeding $10^6$ \cite{Asai:ToeplitzImp:2011}. As discussed above, privacy amplification requires a big block size due to the finite-size key effect \cite{MXF:Finite:2011}, whereas in the application of randomness extraction, a small block size will only reduce the efficiency. Nevertheless, the technique developed in \cite{Asai:ToeplitzImp:2011} could be useful for extractor implementations as well, which we will leave for future investigation.

\section{Randomness Test} \label{Sec:Ext:Test}
\subsection{Statistical tests}
We apply three standard statistical tests---\textsc{diehard},\footnote{www.stat.fsu.edu/pub/diehard/} NIST,\footnote{www.csrc.nist.gov/groups/ST/toolkit/rng/} and \textsc{testu01} \cite{LEcuyer:Testu01:2007}---to evaluate our results. Again, since we only have the data about the QRNG of Ref.~\cite{Xu:QRNG:2012}, we use it as the test of our implementation of randomness extractors. First, the raw data from the QRNG \cite{Xu:QRNG:2012} does not pass the statistical tests due to the classical noises mixed in the raw data and the fact that the as-obtained quantum signals follow a Gaussian distribution instead of a uniform distribution. Secondly, the random numbers from a pseudo-RNG cannot pass all the tests, which exposes its underlying determinism. Finally, we repeatedly operate the Toeplitz-hashing extractor and Trevisan's extractor on our raw data. The outputs from both extractors successfully pass all the standard
statistical tests, which indicates that our postprocessing is effective in extracting out uniform randomness from a weak
randomness source. All the test results are shown in Appendix~\ref{App:Ext:Tests}.

\subsection{Autocorrelation}
\begin{figure*}[htb]
\centering \subfigure[] {\includegraphics [width=6.5cm]
{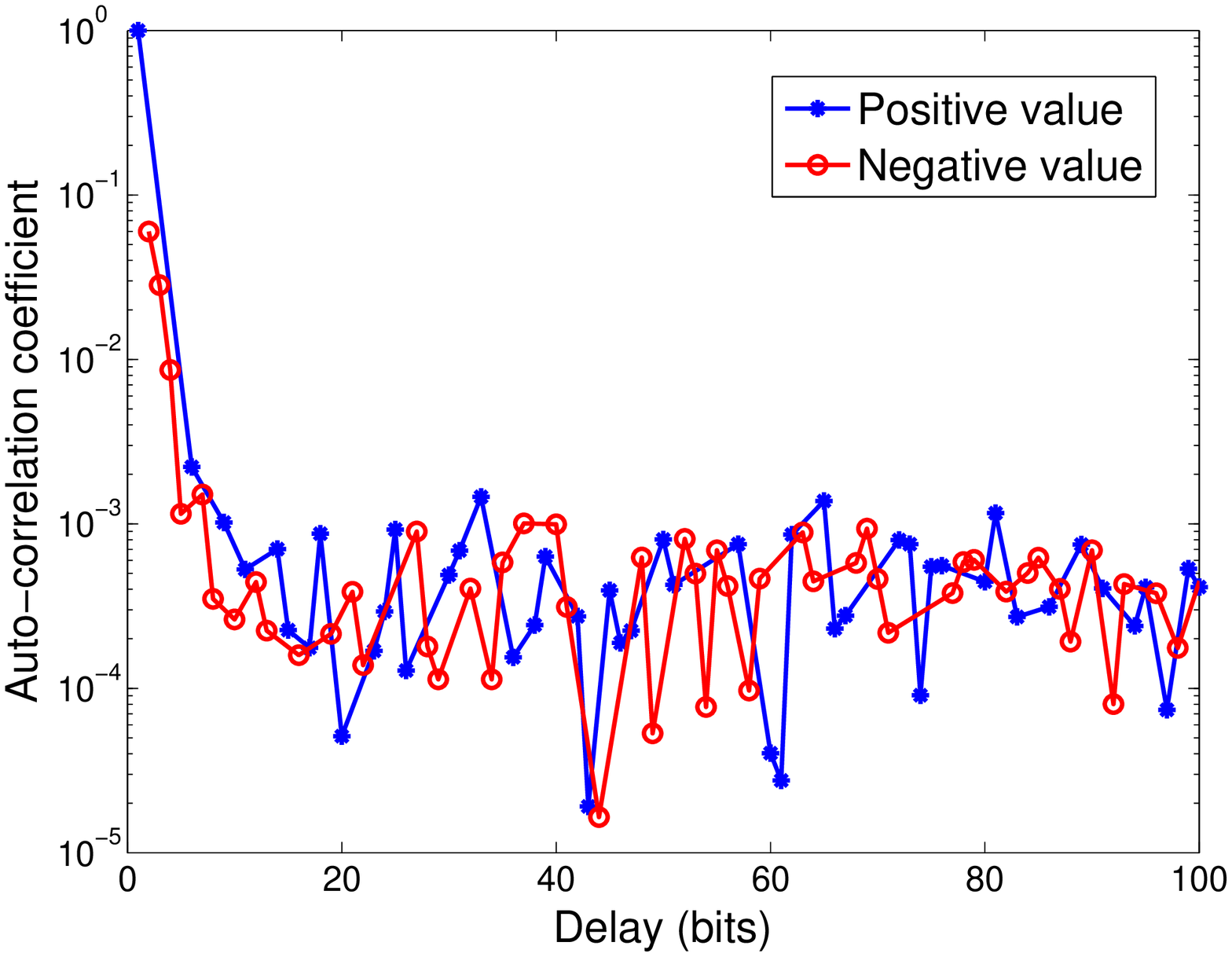} \label{fig:subfig1}} \qquad \subfigure[]
{\includegraphics [width=6.5cm] {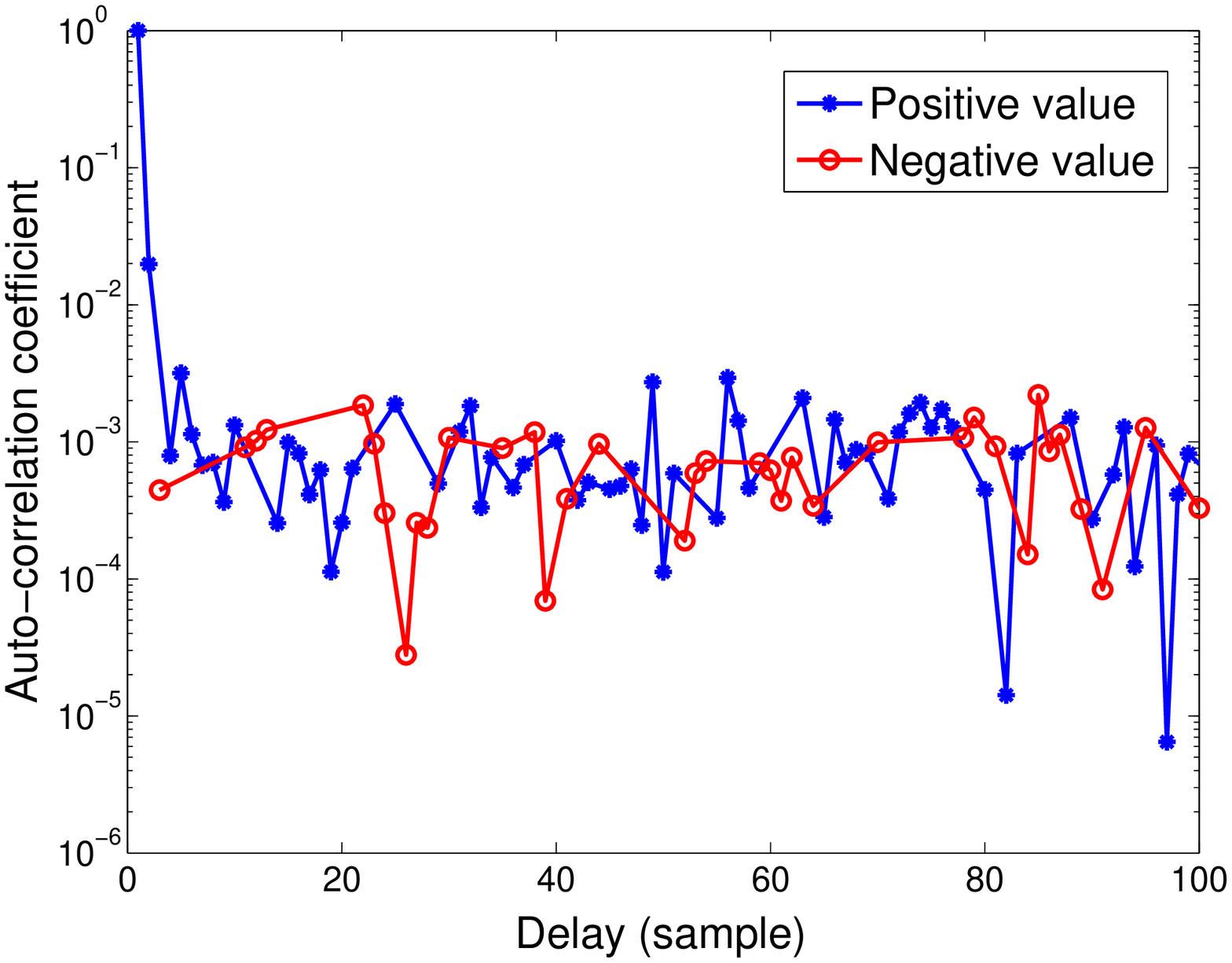}
\label{fig:subfig2}} \qquad \subfigure[] {\includegraphics
[width=6.5cm] {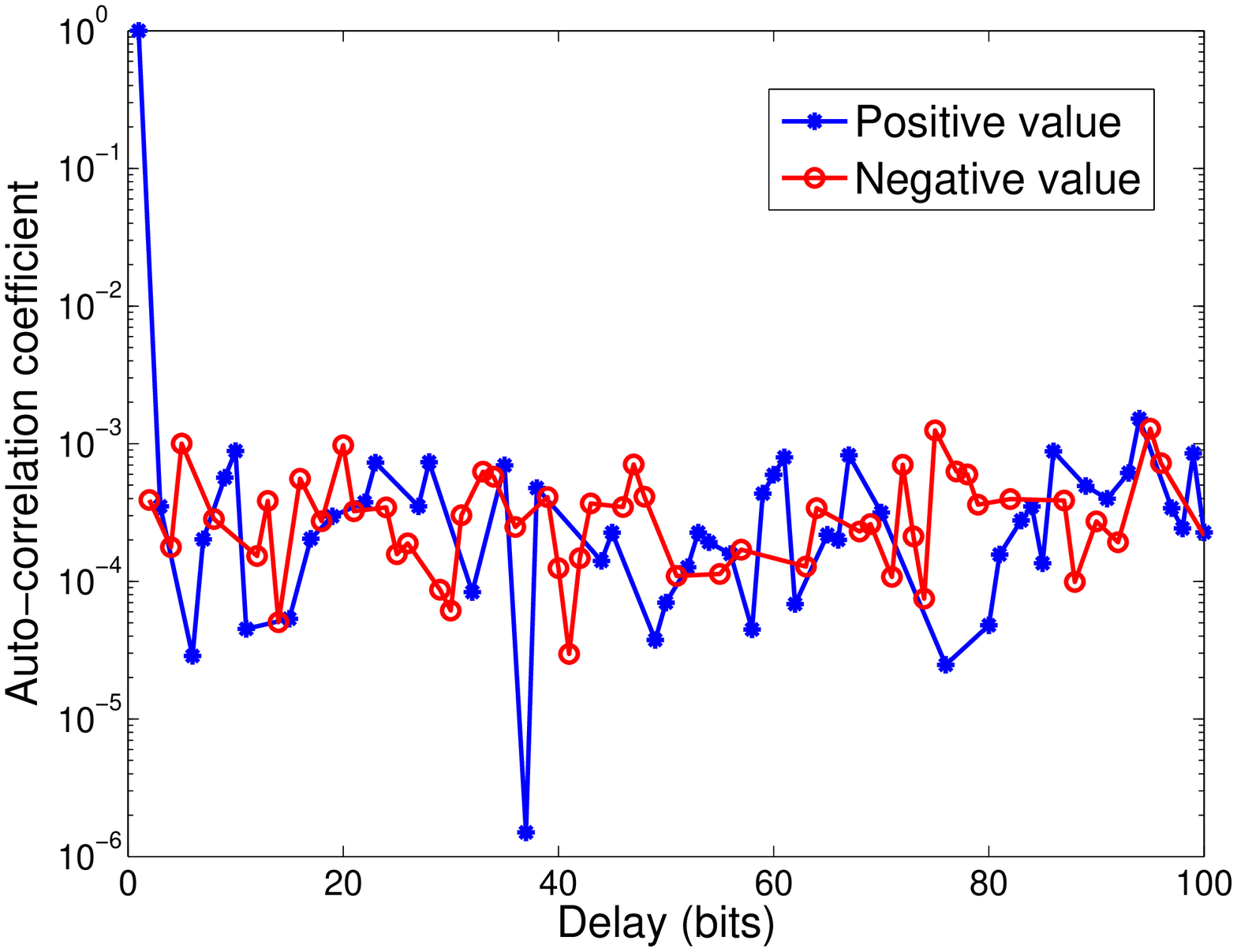} \label{fig:subfig3}} \qquad \subfigure[]
{\includegraphics [width=6.5cm] {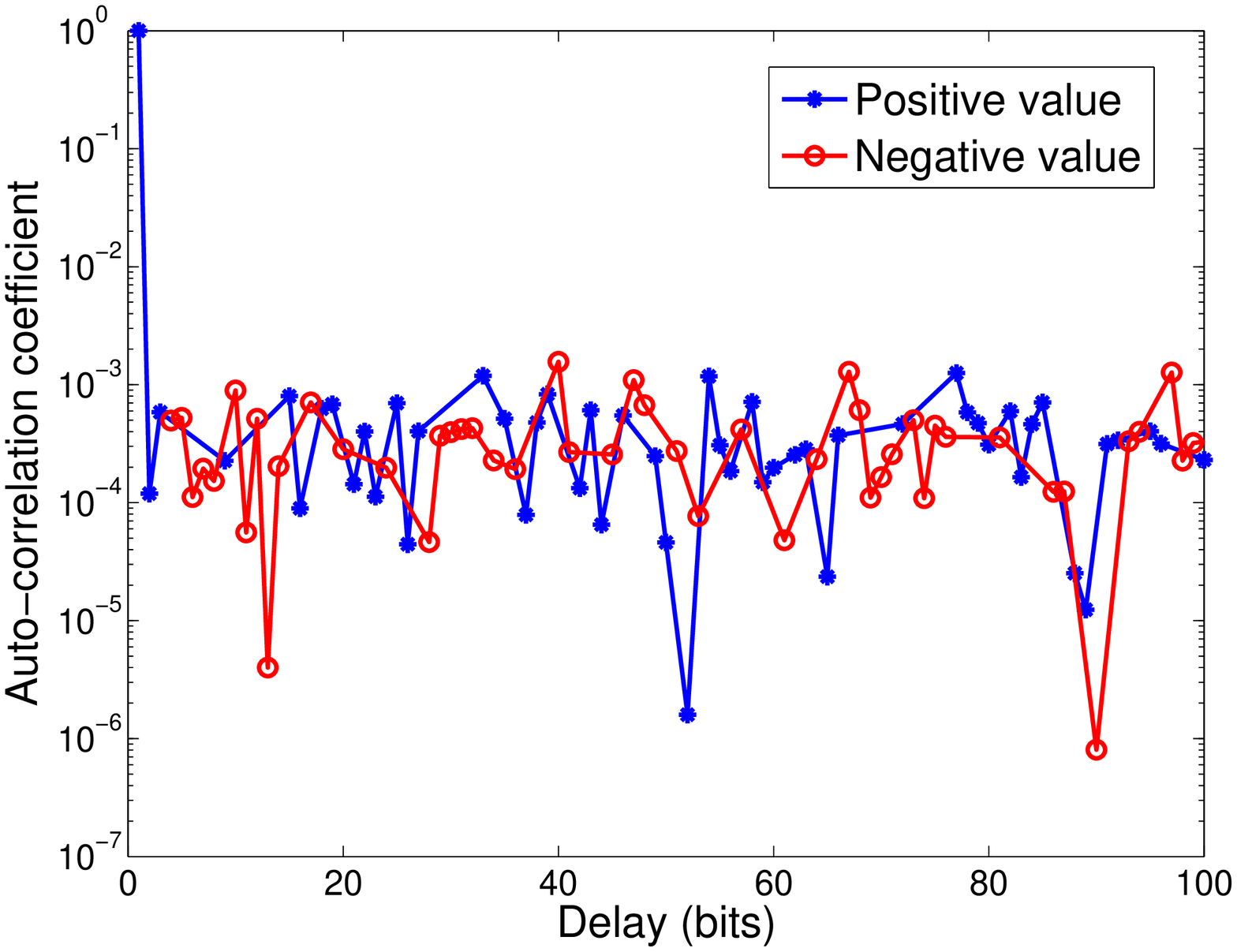} \label{fig:subfig4}}
\caption{ Autocorrelation evaluation results. All normalized correlation is evaluated from a 10 Mb record of the raw data. (a) Autocorrelation of the raw data (between bits). The average value of autocorrelation coefficient is
$9.5 \times 10^{-4}$. The most significant correlations are within 8 bits, due to the usage of 8-bit ADC. (b) Autocorrelation of the raw data (between samples). The average value is $4.9\times10^{-4}$. (c) Autocorrelation of the outcomes from the Toeplitz-hashing extractor. The average value is $-1.0\times10^{-5}$. (d) Autocorrelation of the outcomes from Trevisan's extractor. The average value is $1.6 \times 10^{-5}$. In theory, for a truly random $10\times10^{6}$ bit string, the average normalized correlation coefficient is 0 with a standard deviation of $4\times10^{-4}$.}
\end{figure*}

An alternative approach to verify randomness is evaluating the autocorrelation. The autocorrelations of the raw data are shown in Figs.~\ref{fig:subfig1} (between bits) and \ref{fig:subfig2} (between samples). From Fig.~\ref{fig:subfig1}, we can see that the autocorrelation is significant only within an 8-bit sample, but drop to the vicinity of below $1 \times 10^{-3}$. Also, the low values of the autocorrelation between samples [Fig.~\ref{fig:subfig2}] verify the assumption that the sequence of raw data is i.i.d.~(see Sec.~\ref{Sub:Ext:Model}). We remark that due to the finite bandwidth of a practical detector and statistical fluctuations, the autocorrelation is around $1 \times 10^{-3}$ but never drops to 0.

After postprocessing by either Trevisan's extractor or the Toeplitz-hashing extractor, not only the correlation within 8 bits (from a sample digitalized by an 8-bit ADC) is eliminated, but also the autocorrelation beyond 8 bits drops to $1 \times 10^{-5}$. The autocorrelations of the postprocessing outputs are shown in Figs.~\ref{fig:subfig3} and \ref{fig:subfig4}, where the low residual values indicate the good randomness of our extracted
results.


\section{Concluding remarks} \label{Sec:Eff:Conclusion}
We have modeled QRNG to evaluate the min-entropy of the quantum source, and discussed implementation of a popular extractor---Trevisan's extractor. We have also implemented a Toeplitz-hashing based extractor. We have applied our postprocessing scheme to a recent QRNG implementation \cite{Xu:QRNG:2012} and the min-entropy evaluation procedure on another implementation \cite{Gabriel:QRNG:2010}. The random numbers obtained at the end of postprocessing passed through all the tests of \textsc{diehard}, NIST, and \textsc{testu01}.

From our implementation of Trevisan's extractor, we find that the bottleneck for its extraction speed lies on the one-bit extractor part. Thus, in order to improve the implementation speed, one should investigate the one-bit extractor. In fact, the recent follow-up work by Maurer \emph{et al.}~shows that such improvement can be made \cite{Mauerer:Extraction:2012}. Our prototypical implementation of Trevisan's extractor allows researchers to better understand the complexity and difficulty in the implementation of Trevisan's extractor, thus paving the way to future implementations.

\section{Acknowledgments}
We thank C.-H.~F.~Fung and C.~Rockoff for enlightening discussions. We also thank H.~Zheng and N.~Raghu for the preliminary work on the programming. Support from funding agencies NSERC, the CRC program, CIFAR, MaRS POP and QuantumWorks is gratefully acknowledged. X.~M.~gratefully acknowledges the financial support from the National Basic Research Program of China Grants No.~2011CBA00300 and No.~2011CBA00301; National Natural Science Foundation of China Grants No.~61073174, No.~61033001, No.~61061130540, and No.~61003258; and the 1000 Youth Fellowship program in China.

\begin{appendix}

\section{Trevisan's extractor implementation details} \label{App:Ext:Trevisan}


The choice of block size not only determines the seed cost and security parameter of the random output, but also affects the complexity aspect of the performance. For demonstration purposes, we pick up a set of parameters for Trevisan's extractor, listed in Table~\ref{Tab:Ext:Parameter2}, which can be run sufficiently fast on a personal computer.

\begin{table}[hbt]
\caption{A parameter set for Trevisan's extractor.} \label{Tab:Ext:Parameter2}
\begin{tabular}{ccccc}
\hline
\hline
Extraction efficiency & RS GF($2^{m_e}$) & Design GF($2^{m_d}$) & Input & Output \\
$\rho=1$ & $m_e=128$ & $m_d=8$ & $n_i=2^{15}$ & $n_f=2^{14}$ \\
\hline
Security parameter & ECC codeword & Blocks & Seed &  \\
$\varepsilon=\sqrt{2^{4-m_e} n_in_f^2} $ & $\bar{n}=2^{2m_e}$ & $b=7$ & $d=4m_e^2b$ &  \\
\hline
\hline
\end{tabular}
\end{table}

In this case, the random seed length is larger than the output length, and we can concatenate a hashing-based extractor to make the entropy loss minimum \cite{Raz:Extractor:2002}. We pick up the output length of $n_f=1$ Mb. On one hand, too large a $n_f$ will slow down the extractor, much owing to the $O(n^2)$ complexity with respect to input length; on the other hand, too small a $n_f$ will result in not only high seed cost but also a degradation of security (a larger security parameter $\varepsilon$).


Careful analysis of computational complexity is essential to understanding the tractability or intractability of our implementation given a reasonable computational power. The analysis of complexity of the combinatorial design in Table \ref{Tab:Speed:Design} demonstrates that the most economical parameter in terms of rate is at $n_f=2^{14}$. A smaller parameter will render the design powerless due to associated high key cost, and a larger parameter results in unwieldy complexity growth.

\begin{table}[hbt]
\caption{Real time profile of the speed of combinatoric design. Parameters are selected to result the highest generation rate. Number theoretical operations in $GF_{2^m}$ dominate the speed performance of the ECC, and determines the speed of real-time performance and bit rate (per second).} \label{Tab:Speed:Design}
\begin{tabular}{ccccc}
\hline
\hline
$\log n_f$& Experimental no. of & Theoretical no. of & Experiment no. of $GF_{2^m}$  & Real time\\

& $GF_{2^m}$ operation   & $GF_{2^m}$ operation  & operation per $n_f$ size & (s)\\
\hline

10&65280&262144& 63.75&41.1934\\

11&196352&786432&95.875&124.8\\

12&458496&2097152&111.9375&300.81\\

13&982784&5242880&119.96875&685.91\\

14&203130&12582912&123.984375 & 1603.8\\

15&4128512&29360128&125.99 & 3960.4\\

16&8322816&67108864&126.9960938&10911\\
\hline
\hline
\end{tabular}
\end{table}

As in Table \ref{Tab:Speed:ECC}, the top generation rate of our extractor is 706.8 bits/s; the low speed of the extractor is a consequence of the lack of efficient implementation of finite field operations. Although slow in speed, the results from Trevisan's extractor do pass the statistical tests of \textsc{diehard}. This increase in performance is at the cost of decrease in speed. The severe restriction on speed has limited the usage of Trevisan's extractor in real-time applications.

Our implementation is done on a mere PC, but a mainframe computer can crunch number-theoretical operations much faster than a PC. Furthermore, as a future perspective, once we tackle the implementation on any graphical processing unit (GPU) platforms, the architecture of GPUs will allow us to exploit the intrinsic parallelism of the extractor much more efficiently via multithreading capability.

\begin{table}[hbt]
\caption{Real-time profile of the speed of the error control code (ECC). Parameters are selected to result in the highest generation rate. Number-theoretical operations in $GF_{2^m}$ dominate the speed performance of the ECC, and determine the speed of real-time performance and bit rate (per second).} \label{Tab:Speed:ECC}
\begin{tabular}{cccccc}
\hline
\hline
$n_f$ (power)& Experimental no. of & Theoretical no. of & Experiment no. of $GF_{2^m}$& Real time & Bit rate\\
& $GF_{2^m}$ operation & $GF_{2^m}$ operation & operation per $n_f$ size & (s) & ($s^{-1}$) \\
\hline

1024(10)&15360&16384&15&1.4488&706.8 \\

2048(11)&63488&65536&31&5.9326&345.21121\\

4096(12)&258048&262144&63&23.5451&173.96401 \\

8192(13)&1040384&1048576&127&95.72&173.96 \\

16384(14)&4177920&4194304&255&380.19&43.1 \\

32768(15)&16744482&16777216&511&1536.8&21.32 \\
\hline
\hline
\end{tabular}
\end{table}

\section{Statistical test results} \label{App:Ext:Tests}
We employ three statistical tests---\textsc{diehard}, NIST, and \textsc{testu01} \cite{LEcuyer:Testu01:2007}---to evaluate the randomness of our extracted results from the Toeplitz-hashing extractor and Trevisan's extractor. The test results are shown in Tables~\ref{Tab:Result:dihard}, \ref{Tab:Result:nist}, and \ref{Tab:Result:TestU01}. We can see that, the outputs from two extractors successfully pass all the standard statistic tests. Here, given the constraint of computational power for Trevisan's extractor, we skip the NIST and \textsc{testu01} tests for its results.

Without postprocessing, the raw data cannot pass any statistical tests, which is mainly due to the classical noises mixed in the raw data, and the fact that the measured quantum fluctuations follow Gaussian distribution instead of uniform distribution. This demonstrates the requirement of effective postprocessing in the QRNG.

For control purposes, we also perform the statistic tests on a pseudo-RNG generated from \textsc{matlab2007}. It generates uniformly random numbers from 0 to 255 (as emulation of 8-bit ADC output). The results are shown in Tables \ref{Tab:Result:dihard}, \ref{Tab:Result:nist}, and \ref{Tab:Result:TestU01}. It cannot pass all tests.

\begin{table}[hbt]
\begin{tabular}{lcccccc}
\hline
\hline
& Pseudo-RNG & Raw data & Trevisan's & & Toeplitz hashing & \\
\hline
Statistical test & Result & Result & $P$-value  & Result & $P$-value & Result \\
\hline
  Birthday Spacings [KS]& success& \emph{failure} & 0.822630 & success &0.340863& success\\
  Overlapping permutations &success & \emph{failure}& 0.679927 & success &0.403824& success\\
  Ranks of 31x31 matrices &success &\emph{failure} & 0.419095 & success &0.349441& success\\
  Ranks of 31x32 matrices &success & \emph{failure}& 0.715705 & success &0.816752& success\\
  Ranks of 6x8 matrices [KS]& success& \emph{failure}& 0.195485 & success &0.408573& success\\
  Bit stream test & success & \emph{failure}& 0.048260 & success &0.281680& success\\
  Monkey test OPSO & success& \emph{failure}& 0.027300 & success &0.892600& success\\
  Monkey test OQSO & success & \emph{failure}& 0.023200 & success &0.267200& success\\
  Monkey test DNA & \emph{failure}& \emph{failure}& 0.038000 & success &0.736700& success\\
  Count 1's in stream of bytes & success & \emph{failure}& 0.380162 & success &0.639691&success\\
  Count 1's in specific bytes  &\emph{failure} & \emph{failure}& 0.020417 & success &0.373149&success\\
  Parking lot test [KS] &  \emph{failure}& \emph{failure}& 0.629013  & success &0.151689&success\\
  Minimum distance test [KS]&success & \emph{failure}& 0.019499 & success &0.688780&success\\
  Random spheres test [KS]& success& \emph{failure}& 0.488703 & success &0.939227&success\\
  Squeeze test & success& \emph{failure}& 0.238004 & success &0.155403&success\\
  Overlapping sums test [KS]& success& \emph{failure}& 0.022339  & success &0.909675&success\\
  Runs test (up) [KS]& \emph{failure}& \emph{failure}& 0.403504 & success &0.181024&success\\
  Runs test (down) [KS]& success& \emph{failure}& 0.119132 & success &0.668512&success\\
  Craps test No. of wins & success& \emph{failure}& 0.757521 & success &0.826358&success\\
  Craps test throws/game & success& \emph{failure}& 0.179705 & success &0.862986&success\\
\hline
\hline
\end{tabular}
\caption{\textsc{diehard}. Data size is 240 Mbits. For the cases of multiple $P$-values, a Kolmogorov-Smirnov (KS) test is used to obtain a final $P$-value, which measures the uniformity of the multiple $P$-values. The test is successful if all final $P$-values satisfy $0.01 \leq P \leq 0.99$.} \label{Tab:Result:dihard}
\end{table}

\begin{table}[hbt]
\caption{NIST. Data size is 3.25 Gbits (500 sequences with each sequence around 6.5 Mbits). To pass the test, $P$-value should be larger than the lowest significant level $\alpha=0.01$, and the proportion of sequences satisfying $P > \alpha$ should be greater than 0.976. Where the test has multiple $P$-values, the worst case is selected.} \label{Tab:Result:nist}
\begin{tabular}{lccccc}
\hline
\hline
  & Pseudo-RNG  & Raw data & Toeplitz hashing & & \\
    \hline

      Statistical test & Result & Result & $P$-value& Proportion & Result  \\

    \hline

  Frequency & success& \emph{failure}  & 0.373625 & 0.9900 & success \\
  Block-frequency &success &\emph{failure} & 0.310049 & 0.9960 & success \\
  Cumulative sums &success &\emph{failure}  & 0.422638 & 0.9980 & success \\
  Runs &success &\emph{failure}  & 0.703417 & 0.9900 & success \\
  Longest run & success & \emph{failure}  & 0.013569 & 0.9880 & success \\
  Rank & success & \emph{failure}  & 0.411840 & 0.9940 & success \\
  FFT  & success& \emph{failure}  & 0.987079 & 0.9860 & success \\
  Nonoverlapping template & \emph{failure}  & \emph{failure}  & 0.727851 & 0.9820 & success \\
  Overlapping template & success & \emph{failure}  & 0.110083 & 0.9780 & success \\
  Universal  & success &\emph{failure}  & 0.962688 & 0.9880 & success \\
  Approximate entropy  &success &\emph{failure}  & 0.674543 & 0.9920 & success \\
  Random-excursions  & success &\emph{failure}  & 0.409207 & 0.9900 & success \\
  Random-excursions variant  &success &\emph{failure}  & 0.426358 & 0.9840 & success \\
  Serial  &success &\emph{failure}  & 0.217570 & 0.9860 & success \\
  Linear-complexity  &success &\emph{failure}  & 0.657833 & 0.9940 & success \\
\hline
\hline
\end{tabular}
\end{table}

\begin{table}[hbt]
\caption{\textsc{testu01} (small crush). Given the constraint of the data size and computational power of crush and big crush of \textsc{testu01}, we only perform the small crush test here. Data size is 8 Gbits. The $P$-value from a failing test converges to 0 or 1. Where the test has multiple $P$-values, the worst case is selected.} \label{Tab:Result:TestU01}
\begin{tabular}{lcccc}
\hline
\hline
     & Pseudo-RNG & Raw data & Toeplitz-hashing & \\
  \hline
        Statistical test & Result & Result & $P$-value & Result \\

   \hline
  BirthdaySpacings & Success &\emph{failure}  & 0.5300 & success \\
  Collision & Success&\emph{failure}  & 0.1500 & success \\
  Gap Chi-square & success &\emph{failure}  & 0.8900 & success \\
  SimpPoker Chi-square & success &\emph{failure}  & 0.3500 & success \\
  CouponCollector Chi-square & success &\emph{failure}  & 0.6700 & success \\
  MaxOft Chi-square &success &\emph{failure}  & 0.6900 & success \\
  MaxOft Anderson-Darling & success &\emph{failure}  & 0.9500 & success \\
  WeightDistrib Chi-square & success &\emph{failure}  & 0.5600 & success \\
  MatrixRank Chi-square & success &\emph{failure}  & 0.5100 & success \\
  Hammingindep Chi-square & success &\emph{failure}  & 0.1000 & success \\
  RandomWalk1 H Chi-square & success &\emph{failure}  & 0.9931 & success \\
  RandomWalk1 M Chi-square &success &\emph{failure}  & 0.8300 & success \\
  RandomWalk1 J Chi-square & success &\emph{failure}  & 0.9400 & success \\
  RandomWalk1 R Chi-square & success &\emph{failure}  & 0.7000 & success \\
  RandomWalk1 C Chi-square & success &\emph{failure}  & 0.6600 & success \\
\hline
\hline
\end{tabular}
\end{table}

\end{appendix}

\bibliographystyle{apsrev4-1}

\bibliography{Bibli}


\end{document}